\documentclass[11pt]{article}

\date{}

\usepackage{tabularx}
\usepackage{gensymb}
\usepackage{textcomp}
\usepackage{adjustbox}
\usepackage[utf8]{inputenc} 
\usepackage[T1]{fontenc}    
\usepackage{url}            
\usepackage{booktabs}       
\usepackage{amsfonts}       
\usepackage{nicefrac}       
\usepackage{microtype,nicefrac}      
\usepackage{xspace}
\usepackage{natbib}
\usepackage{comment}
\usepackage{graphicx}
\usepackage{rotating}
\usepackage{amssymb}
\usepackage{autobreak}
\usepackage{mathtools}
\usepackage[dvipsnames]{xcolor}
\usepackage{bbm}
\usepackage[ruled,vlined, linesnumbered]{algorithm2e}
\usepackage{algorithmic}
\usepackage[parfill]{parskip}
\usepackage{color}
\usepackage{amsmath} 
\usepackage{graphicx}
\usepackage{subfig}
\usepackage{xr}
\usepackage[urlcolor=blue,citecolor=black,linkcolor=black]{hyperref}
\usepackage[blocks]{authblk}
\allowdisplaybreaks

\title{Enforcing Reciprocity in Operator Learning \\for Seismic Wave Propagation}

\author[1]{Caifeng Zou\thanks{Corresponding author: czou@caltech.edu}}
\author[2]{Yaozhong Shi}
\author[1]{Zachary E. Ross}
\author[1]{Robert W. Clayton}
\author[3]{Kamyar Azizzadenesheli}

\affil[1]{Seismological Laboratory, California Institute of Technology, Pasadena, California, U.S.A.}
\affil[2]{Division of Engineering and Applied Sciences, California Institute of Technology, Pasadena, California, U.S.A.}
\affil[3]{Nvidia Corporation, Santa Clara, California, U.S.A.}

\begin{document}

\label{firstpage}

\maketitle

\begin{abstract}
Accurate and efficient wavefield modeling underpins seismic structure and source studies. Traditional methods comply with physical laws but are computationally intensive. Data-driven methods, while opening new avenues for advancement, have yet to incorporate strict physical consistency. The principle of reciprocity is one of the most fundamental physical laws in wave propagation. We introduce the Reciprocity-Enforced Neural Operator (RENO), a transformer-based architecture for modeling seismic wave propagation that hard-codes the reciprocity principle. The model leverages the cross-attention mechanism and commutative operations to guarantee invariance under swapping source and receiver positions. Beyond improved physical consistency, the proposed architecture supports simultaneous realizations for multiple sources. This yields an order-of-magnitude inference speedup at a similar memory footprint over a conventional neural operator on a realistic multi-source configuration. We demonstrate the functionality using the reciprocity relation for particle velocity fields under single forces. This architecture is also applicable to pressure fields under dilatational sources and travel-time fields governed by the eikonal equation, paving the way for encoding more complex reciprocity relations.
\end{abstract}

\section{Introduction}
Accurate and efficient modeling of seismic wave propagation is essential for subsurface structure imaging, ground motion simulation, and earthquake source inversion. Traditional numerical solvers, such as finite difference and finite element methods, have theoretical guidance for modeling accuracy and stability (e.g., the number of grid points required per wavelength). However, computational cost is often the primary bottleneck, particularly for large-scale inverse problems or applications requiring extensive parameter sweeps. The emergence of machine learning has offered promising pathways for accelerating seismic wavefield modeling.

Machine learning methods for solving partial differential equations (PDEs) are currently dominated by two lines of work:  Physics-Informed Neural Networks (PINNs) \citep{raissi2019physics} and Neural Operators \citep{li2020neural}. The former approximates the solution function, while the latter learns the solution operator. Training a PINN is challenging, because it uses the PDE as a soft constraint, resulting in a highly non-convex loss landscape. Moreover, it only approximates the solution function to a single instance of the PDE family and requires retraining for different PDE parameters. Consequently, PINN applications in seismology are often constrained to a specific velocity model \citep{Alkhalifah,huang2022pinnup,huang2022single,Rasht‐Behesht,ren2024seismicnet,Song2021,Song2022}. Neural operators do not have these limitations, because they learn maps between function spaces from data (which could include simulations). They exhibit excellent generalization performance in modeling seismic wave propagation \citep{huang2025learned,Kong2025,lehmann20243d,lehmann2025multiple,yang2021seismic,yang2023rapid,zou2024deep,zou2025ambient}. However, neural operators are primarily data-driven (despite physics-informed variants \citep{li2024physics}) and often lack the precision of traditional solvers. Furthermore, they may violate physical laws if not explicitly constrained.

The reciprocity principle is a fundamental physical law in wave propagation that relates two seismic wavefields in a medium under interchange of source and receiver positions \citep{arntsen2000new,Knopoff1959}. This principle is most commonly used for efficient wavefield simulations where sources significantly outnumber receivers \citep{eisner2001reciprocity,graves1992modeling}, thereby facilitating source and structure inversion \citep{petrov2019estimation,zhao2006strain}. Although its concise formula and ease of implementation cast reciprocity as an ideal constraint for training neural surrogates, this is sparsely documented. Only recently have \cite{geng2025seismic} and \cite{wang2026reciprocity} employed reciprocity as a soft constraint in training PINNs for travel-time simulation. However, hard-coding this physical symmetry into model architectures remains unexplored.

Here, we introduce the Reciprocity-Enforced Neural Operator (RENO), a transformer-based architecture that hard-codes the reciprocity principle into seismic wavefield modeling. We start by introducing the reciprocity relations addressed in this study and basics of neural operators. We then detail the RENO architecture. Following a description of the synthetic data set used for training, we conduct experiments to compare RENO with a conventional neural operator in terms of data efficiency, physical consistency, and computational efficiency. We also demonstrate RENO's capability to perform rapid full waveform inversion via automatic differentiation. Finally, we discuss the theoretical and practical distinctions between hard and soft constraints in physics-informed machine learning and summarize the advantages, applications, and limitations of the proposed framework.

\section{Preliminaries}
\subsection{Reciprocity relations}
\begin{figure}
\centering
\includegraphics[width=0.8\textwidth]{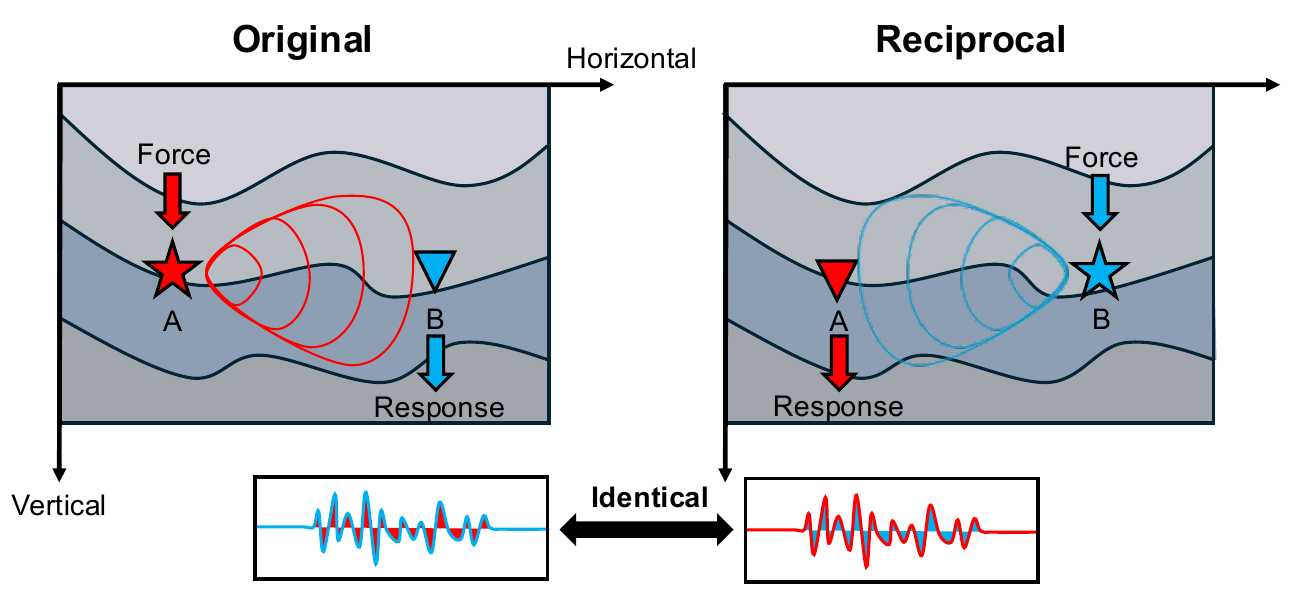}
\caption{Schematic diagrams of the reciprocity relation for single vertical forces at the sources and vertical particle velocity responses at the receivers. The diagram on the left illustrates the original experiment, where the source is placed at point A (red star) and the receiver at point B (blue triangle). The diagram on the right shows the reciprocal experiment, where the source is relocated to point B (blue star) and the receiver to point A (red triangle). Both experiments yield identical waveforms at the receivers.}
\label{schematic}
\end{figure}

The reciprocity principle describes a certain invariance in a medium where the source and receiver positions are interchanged. The reciprocity relations are categorized by types of sources (e.g., single forces and force couples) and physical properties recorded at receivers (e.g., particle velocity, strain, and stress) \citep{arntsen2000new}. In this study, we take the example of single forces and particle velocity fields, the reciprocity relation for which holds in inhomogeneous, anisotropic, viscoelastic media is
\begin{equation}
\begin{aligned} 
v_n^{m}(\mathbf{x}_0', t; \mathbf{x}_0)
=
v_m^{n}(\mathbf{x}_0, t; \mathbf{x}_0'),
\label{treci}
\end{aligned}
\end{equation}
where $v_n^{m}(\mathbf{x}_0', t; \mathbf{x}_0)$ indicates the $n$ component of the particle velocity at position $\mathbf{x}_0'$ due to a body force acting in the $m$ direction at position $\mathbf{x}_0$ as a function of time $t$. In the frequency domain, the relation becomes
\begin{equation}
\begin{aligned} 
V_n^{m}(\mathbf{x}_0', \omega; \mathbf{x}_0)
=
V_m^{n}(\mathbf{x}_0, \omega; \mathbf{x}_0'),
\label{wreci}
\end{aligned}
\end{equation}
where $V$ is the Fourier transform of $v$ and $\omega$ is the angular frequency. To further simplify the problem, we only consider $m=n$ so that the directional force can be fixed. We focus on single vertical forces and vertical velocity fields, as the horizontal cases should be the same. Thus, the target variable is an invariant scalar when swapping the source and receiver. This configuration applies to particle velocity fields under single forces in the same direction, pressure fields under dilatational sources, and travel-time fields governed by the eikonal equation. Figure \ref{schematic} is a schematic for the reciprocity relation studied here. Future studies involving vectors or different types of sources and responses will be extensions of this pilot study.

\subsection{Neural operators}
Neural Operators are models for learning maps between function spaces, making them a rapid, powerful tool for solving PDEs \citep{azizzadenesheli2024neural,li2020neural}. What distinguishes neural operators from classical neural networks is their discretization agnosticism: a model is not tied to a specific discretization or geometry. While Physics-Informed Neural Networks (PINNs) \citep{raissi2019physics} are also discretization-agnostic, they approximate the solution function for a specific instance. In contrast, neural operators are trained to learn a family of PDEs with varying parameters. Since their inception, this class of machine learning models has seen many variants. For example, the Fourier Neural Operator (FNO) speeds up computations via spectral-domain convolutions \citep{li2020fourier}. The Geometry-Informed Neural Operator (GINO) integrates the ability of graph neural operators (GNOs) to handle irregular grids with the computational efficiency of FNOs \citep{li2023geometry}. The Physics-Informed Neural Operator (PINO) incorporates the function optimization capacity of PINNs into operator learning, combining advantages of both worlds  \citep{li2024physics,ma2026effective}. More recent variants integrate transformers \citep{vaswani2017attention}, which have demonstrated transformative success across various fields, to boost the performance and scalability of neural operators \citep{alkin2024universal,shi2025mesh,wu2024transolver}.

\section{Methods}
\begin{figure}
\centering
\includegraphics[width=0.8\textwidth]{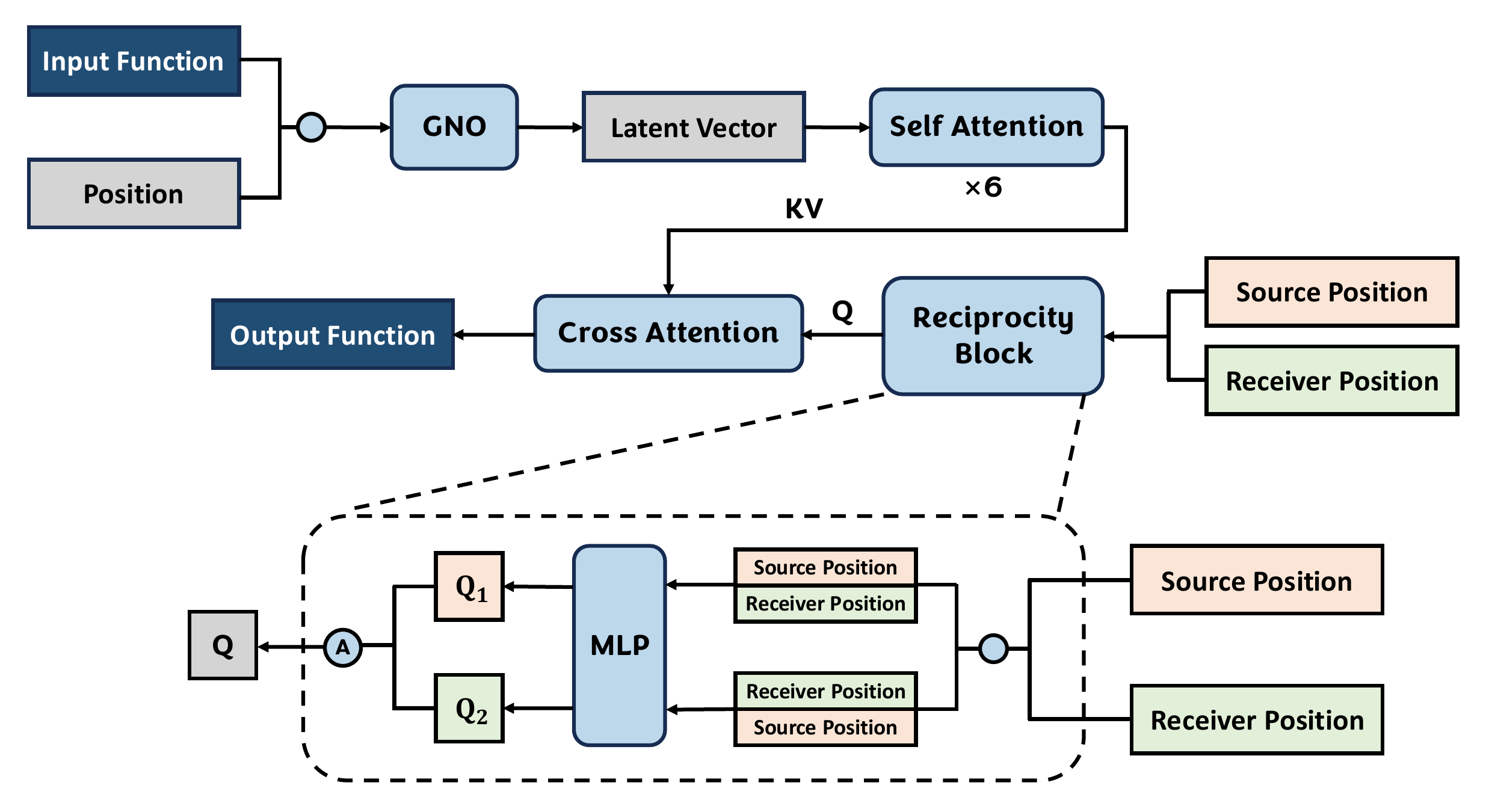}
\caption{RENO architecture. Blue circles denote concatenation. $A$ in the blue circle denotes an averaging operation. $Q$ represents queries and $KV$ represents key-value pairs. All positions are encoded using sinusoidal embeddings.}
\label{architecture}
\end{figure}

In this study, we develop a neural operator architecture that explicitly enforces reciprocity, referred to as RENO (Reciprocity-Enforced Neural Operator). Inspired by the Mesh-Informed Neural Operator (MINO) for generative modeling \citep{shi2025mesh}, we adapt its architecture to ensure reciprocal symmetry. Figure \ref{architecture} outlines the RENO architecture. The input function is represented as a point cloud of values at discrete positions. Since the number of points (tokens) is typically too large for transformers to handle directly due to their quadratic complexity, a GNO is used to compress the input into a latent function represented by a small number of ``super nodes''. This function can then be efficiently processed by self-attention layers. The GNO acts as a discretization-agnostic encoder, ensuring the architecture generalizes to arbitrary input geometries. It also enables RENO to achieve linear complexity with respect to the number of mesh points. RENO uses an encoder-decoder structure to map an input function $\mathbf{f}$ and position $\mathbf{x}$ to an output function $\mathbf{y}$ through three main parts:
\begin{equation}
\begin{aligned} 
\textbf{Encoder:} \quad KV = \text{SelfAttention}^6(\text{GNO}(\mathbf{f}, \mathbf{x}))
\label{encoder}
\end{aligned}
\end{equation}

\begin{equation}
\begin{aligned} 
\textbf{Reciprocity Block:} \quad Q = \text{ReciprocityBlock}(\mathbf{x}_s, \mathbf{x}_r)
\label{reciblock}
\end{aligned}
\end{equation}

\begin{equation}
\begin{aligned} 
\textbf{Decoder:} \quad \mathbf{y} = \text{CrossAttention}(Q, KV)
\label{decoder}
\end{aligned}
\end{equation}

We highlight the Reciprocity Block, where reciprocity is made a hard constraint via a series of commutative operations. Specifically, the source and receiver positions, $\mathbf{x}_s$ and $\mathbf{x}_r$, are concatenated in both permutations and passed through the same Multi-Layer Perceptron (MLP):
\begin{equation}
\begin{aligned} 
Q_1 = \text{MLP}(\mathbf{x}_s, \mathbf{x}_r), \quad\ Q_2 = \text{MLP}(\mathbf{x}_r, \mathbf{x}_s).
\label{mlp}
\end{aligned}
\end{equation}
The MLP consists of two layers with $256$ hidden neurons, using a GELU \citep{hendrycks2016gaussian} activation function following the first layer. The resulting output vectors, $Q_1$ and $Q_2$, are averaged to produce the final query:
\begin{equation}
\begin{aligned} 
Q = \frac{Q_1 + Q_2}{2}.
\label{avr}
\end{aligned}
\end{equation}
This query is then input to the cross-attention decoder to obtain solutions for the corresponding source-receiver pairs. Regarding concepts such as $Q$ (query), $KV$ (key-value), self- and cross- attention or more technical details, we refer the interested readers to \cite{vaswani2017attention} and \cite{shi2025mesh}.

To avoid complexity arising from the time dimension, we train a Helmholtz neural operator for frequency-domain solutions \citep{zou2024deep}, so that both the input and output functions are two-dimensional in space. The input includes $V_P$, $V_S$, and the frequency value parameterized by a constant function. The output is the frequency-domain particle velocity solution for the queried source-receiver pair. The time-domain solution can be obtained via inverse Fourier transform.

\section{Experiments}
In this section, we compare the RENO with a conventional model that is not explicitly informed of reciprocity during training; rather this conventional model learns reciprocity implicitly by learning to solve the PDE. We remove the source-receiver symmetry in the model architecture and make the source location as an input feature along with velocity and frequency information \citep{zou2024deep,zou2025ambient}. We query the waveform solutions at every grid point on the free surface for both models. We use synthetic data generated with a staggered grid finite difference solver \citep{richardson_alan_2025} to train the neural operators. The solver applies to the isotropic elastic wave equation in the particle velocity-stress formulation. The elastic medium is parameterized by $V_P$, $V_S$, and density, of which $V_S$ models are generated from two-dimensional Matérn random fields superimposed on a one-dimensional background model, and the corresponding $V_P$ and density models are derived from $V_S$ using empirical relations \citep{brocher2005empirical}. Each simulation has a single source, the location of which is uniformly sampled from the free surface. The source is a vertical force and the source time function is a Ricker wavelet with a central frequency of $0.3$ Hz. The source time function is fixed and is not treated as an input variable to the neural operator in this study. In practice, one would likely preprocess the data with filtering and deconvolution to make the sources approximately equal. The simulated wavefield lasts $50$ s using a time step of $0.001$ s, which is Fourier-transformed and filtered to $[0.1,0.5]$ Hz to obtain frequency-domain solutions for training the Helmholtz neural operator. This computational overhead can be avoided if a Helmholtz solver is used \citep{chen2013optimal,huang2021finite}. The computational domain is $85$ km (horizontal) $\times$ $20$ km (vertical) with $125$ m spacing. For training, the medium parameters are downsampled by a factor of $2$ for computational efficiency, resulting in a $339 \times 81$ grid. The neural operator is trained to solve for the vertical particle velocity field, for which the reciprocity relation holds under vertical forces.

\subsection{Reciprocal generalization}
\begin{figure}
\centering
\includegraphics[width=0.8\textwidth]{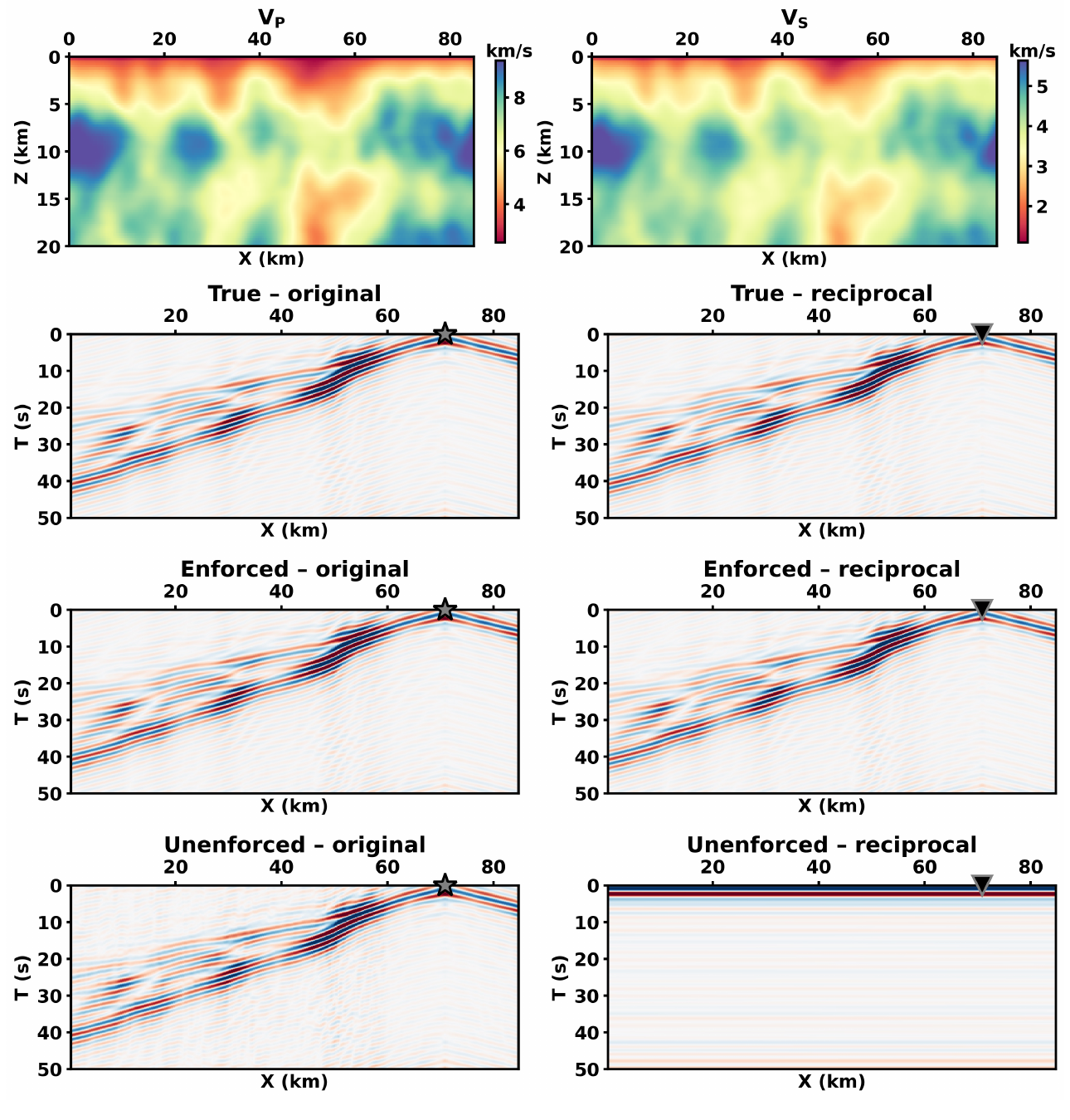}
\caption{Reciprocal experiments where sources and receivers are swapped. The star marks the source, which becomes the receiver marked with a triangle in the reciprocal experiment. The first row shows the input velocity models. The second row shows the waveform solutions from the finite difference solver, taken as ground truth. The third and fourth rows show the predictions from the reciprocity-enforced and unenforced neural operators, respectively. Both models are trained on a single simulation (the original one).}
\label{fit1}
\end{figure}

We start with demonstrating the fundamental difference in how the reciprocity-enforced and unenforced models internalize physical laws. For the purpose of this basic demonstration, we fit both models to just a single simulation and evaluate them on the reciprocal solutions, where the source becomes the receiver and the receivers become sources. Figure \ref{fit1} shows the results. Based on the reciprocity principle, the original and reciprocal solutions should be identical, as verified by the finite difference simulations (ground truth). The RENO, trained to accurately fit the original simulation, generalizes to the reciprocal case with identical accuracy without prior exposure to such data, owing to the imposed hard-coded constraint. However, the conventional operator without this constraint fails to generate any meaningful seismic signal in the reciprocal test, because it has not learned sufficient physics from a single simulation to generalize. This example suggests that, by enforcing reciprocity as an architectural constraint, the model can leverage physical symmetry to map unseen reciprocal pairs, effectively doubling the information underlying each training instance. From another perspective, if both directions of every reciprocal pair were provided (which is computationally prohibitive), RENO would presumably converge approximately twice as fast as a conventional model.

\subsection{Learning behavior}
\begin{figure}
\centering
\includegraphics[width=1.0\textwidth]{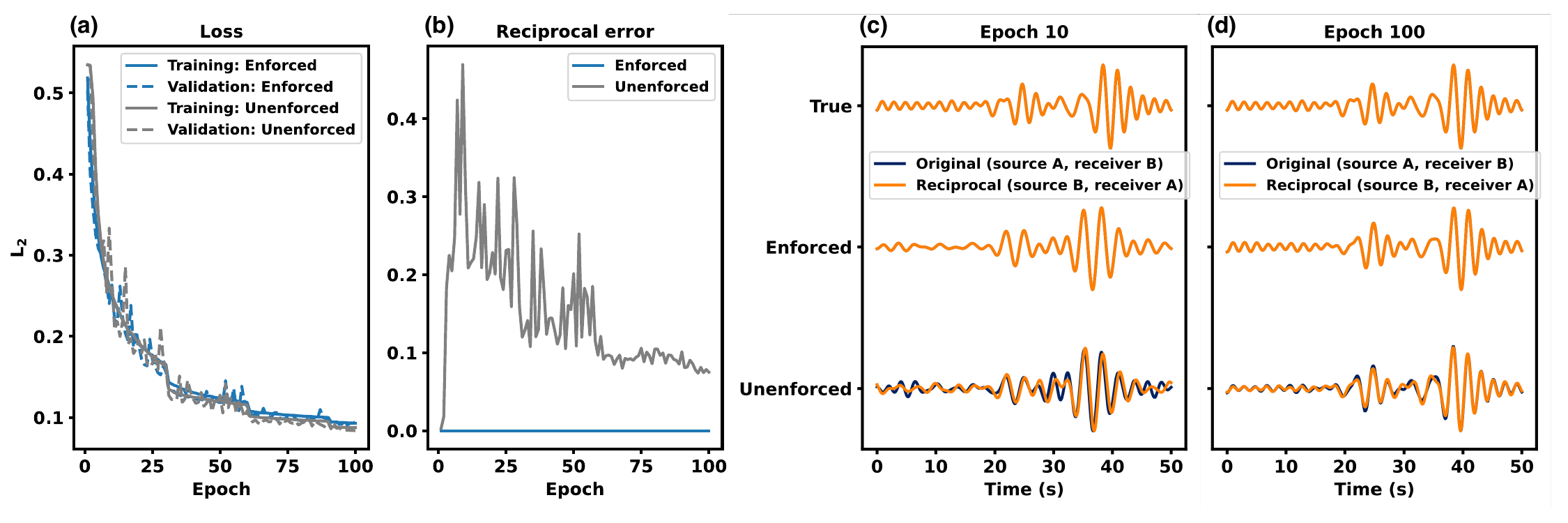}
\caption{(a) Loss and (b) reciprocal error curves during training for the reciprocity-enforced and unenforced models. Validation examples of reciprocal waveforms (the source and receiver positions of which are swapped, as illustrated in Figure \ref{schematic}) at (c) epoch 10 and (d) epoch 100.}
\label{train_curves}
\end{figure}

Here, we study how the two models learn the physics when trained on a reasonable amount of data. We use a normalized $\ell_2$-norm as the loss function. A total of $10000$ simulations ($10000$ different velocity models) are used for training and $1000$ simulations are used for validation. We define the reciprocal error as 
\begin{equation}
\begin{aligned} 
\text{Reciprocal error}=\frac{\left\| \mathcal{L}\left(\mathbf{x}_s,\mathbf{x}_r,*\right)-\mathcal{L}\left(\mathbf{x}_r,\mathbf{x}_s,*\right)\right\|_{2}}{\left\| \mathcal{L}\left(\mathbf{x}_s,\mathbf{x}_r,*\right)\right\|_{2}},
\label{ereci}
\end{aligned}
\end{equation}
where $\mathcal{L}$ denotes the learned neural operator, $\mathbf{x}_s$ and $\mathbf{x}_r$ denote the source and receiver positions, respectively, and $*$ represents other input to the operator. Ideally, this error should be zero. For computational efficiency, a random source-receiver pair is selected from the validation set at each iteration to compute the reciprocal error. Figure \ref{train_curves}a plots the $\ell_2$ loss during training for the reciprocity-enforced and unenforced models, shown in blue and gray, respectively. This metric is computed over solutions at all receivers for a single source and thus does not honor the reciprocity principle but reflects averaged overall performance. In terms of the $\ell_2$ loss, both models converge to similar accuracy, while the RENO performs slightly better at earlier epochs. However, since reciprocity is not hard-coded in the conventional model, it can only acquire this property indirectly from data. Note that the training set does not explicitly include reciprocal pairs (each velocity model is associated with a single source), so the conventional model can only learn the reciprocity principle as part of the general wave physics present in the data. While it might asymptotically conform to the law of reciprocity given infinite data and training, this is not guaranteed in practice. As shown in Figure \ref{train_curves}b, the reciprocal error of the conventional model generally decreases during training but does not ultimately reach zero. It is only zero at the beginning due to initialization. The RENO maintains zero reciprocal error, as designed. Figure \ref{train_curves}c shows an example pair of reciprocal waveforms (the source and receiver positions of which are swapped) from validation data at an early stage of training. The pair of waveforms should be identical according to the reciprocity principle, as illustrated in Figure \ref{schematic}. Although RENO predicts a phase advance relative to the ground truth, it strictly follows the law of reciprocity. In contrast, the conventional model violates reciprocity and yields less accurate predictions. By the end of training, RENO predicts the waveform pair both accurately and identically, as shown in Figure \ref{train_curves}d. While the conventional model improves in both predictive accuracy and adherence to reciprocity, it still fails to achieve strict symmetry. In summary, the RENO has the physical law hard-coded into its architecture, whereas a conventional model must learn it from scratch. Although both models achieve comparable overall accuracy given proper training, the physics-embedded model is expected to be more robust to changes in the experimental setup. Above all, it is a perfect model in terms of the reciprocity metric.

\subsection{Computational efficiency}
The source-receiver symmetry in the RENO architecture not only guarantees adherence to this physical principle but also enables computing solutions for multiple sources in a single query. Unlike conventional operators or numerical solvers restricted to single-source formulations, this parallelization occurs within the query mechanism of the transformer backbone rather than at the batch level, thus significantly improving computational efficiency. Since there is no correlation between queried source-receiver pairs, arbitrarily many sources can be solved for simultaneously, provided there is sufficient memory. RENO's computational complexity scales linearly with both the number of source-receiver pairs and the mesh density.  

We demonstrate this computational advantage using a community velocity model \citep{lee2014full} configured with $234$ sources and $339$ receivers on the free surface, mimicking a real seismic dataset \citep{BASIN2018}. The RENO computes the Helmholtz solutions for all $234 \times 339$ source-receiver pairs in a single query, while the conventional model requires increasing the batch size to parallelize across sources. Using an NVIDIA RTX A6000 GPU with a memory capacity of $46$ GB, the RENO can complete the computation in a single batch, whereas the conventional model encounters an out-of-memory issue and must iterate over multiple batches. At a comparable memory footprint, the RENO exhibits a $30$-fold speedup over the conventional model during inference. Table \ref{efficiency} summarizes the runtime and GPU memory usage, alongside the model size and training cost. The speedup claimed here is based on a controlled, fair comparison between the two models, as we have intentionally kept all the other configurations (hardware, model size, training cost, memory usage) at comparable scales. This speedup is transferable to both inversion and training, provided that the training set includes reciprocal pairs. Note that the comparison here is made between a RENO and a conventional neural operator. The latter has been shown to be about two orders of magnitude faster than traditional numerical solvers like the spectral element method \citep{zou2024deep}. 

\begin{table}
\centering
\caption{Computational efficiency and GPU memory usage for the reciprocity-enforced and unenforced models during inference. The model size and training cost are also provided. The comparison is made on a community velocity model with $234$ sources and $339$ receivers on the free surface, using an NVIDIA RTX A6000 GPU.}
\label{efficiency}
\small
\newcolumntype{C}{>{\centering\arraybackslash}X}

\begin{tabularx}{\textwidth}{@{}lCCCCC@{}}
\toprule
\textbf{Model} & \textbf{Number of Parameters} & \textbf{Training Time} & \textbf{Training GPU Usage} & \textbf{Inference Time} & \textbf{Inference GPU Usage} \\
\midrule
Reciprocity-Enforced & 892290 & 32 h & 10578 MB & 0.31 s & 18182 MB \\ \addlinespace
Unenforced & 859778& 34 h & 10576 MB & 9.34 s & 18966 MB \\
\bottomrule
\end{tabularx}
\end{table}

\subsection{Full waveform inversion}
\begin{figure}
\centering
\includegraphics[width=0.8\textwidth]{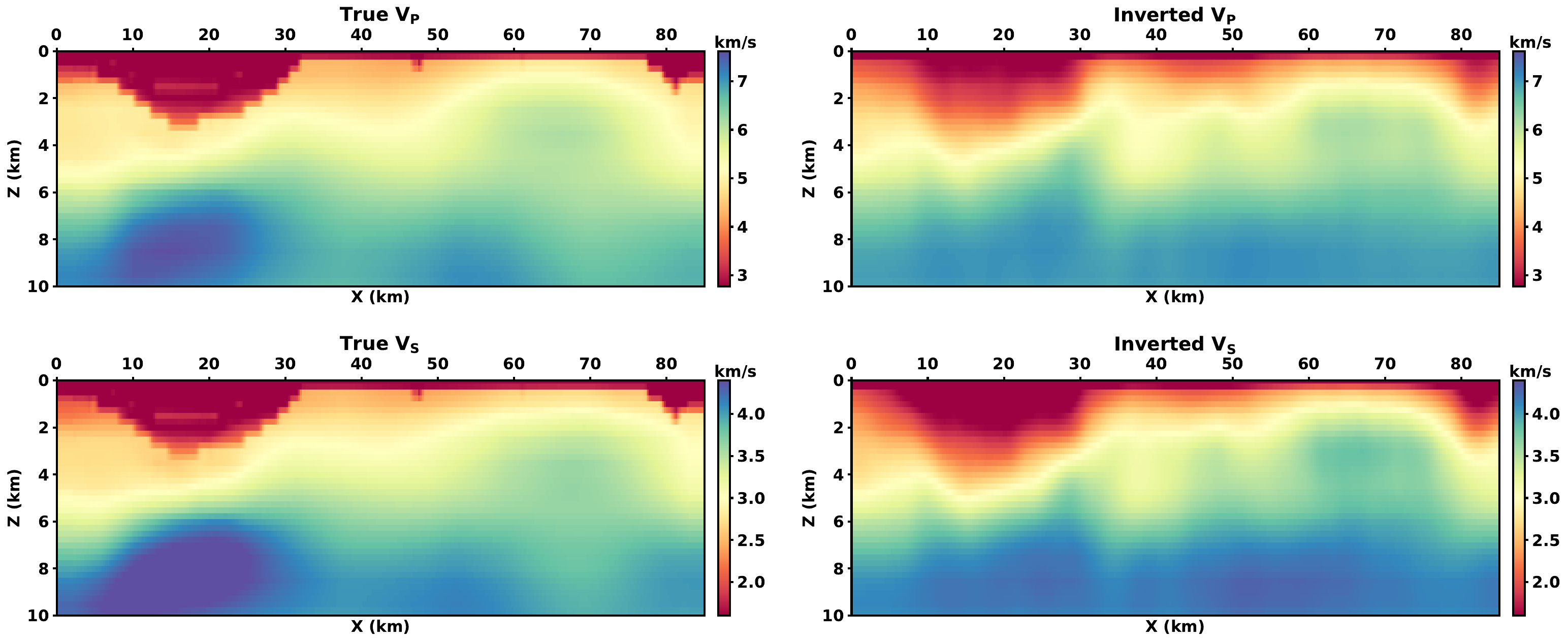}
\caption{Full waveform inversion results for a community velocity model with $234$ sources and $339$ receivers on the free surface, using the trained RENO with automatic differentiation.}
\label{inv}
\end{figure}

Neural operators can be used with automatic differentiation for full waveform inversion \citep{yang2023rapid,zou2025ambient}. Using the trained RENO as the forward modeling engine, we conduct an inversion experiment on the same velocity model (unseen in training) and source-receiver configuration as in the previous section. We average the ground truth horizontally to obtain a one-dimensional model used as the initial model for inversion. Figure \ref{inv} shows the inversion results, where the structural features of sedimentary basins are recovered successfully in both $V_P$ and $V_S$. The inversion runs for $40$ epochs with a batch size of one, taking $27.25$ seconds and consuming $2083$ MB GPU memory. The rapid inversion process benefits from RENO's capability to simultaneously process multiple sources at the query level. In contrast, given the same GPU memory allocation, the conventional model takes $750.96$ seconds to complete the same experiment.

\section{Discussion and Conclusions }
In physics-informed machine learning, there are two primary directions: hard constraints and soft constraints. Hard constraints involve designing a model architecture whose output strictly obeys physical laws, regardless of training. Conversely, soft constraints typically involve a regularization term to be minimized during training, with a classic example being PINNs \citep{raissi2019physics}. These methods differ in several key aspects. For example, models with hard constraints are often more efficient to train, because computing soft-constraint regularization terms may require extra model calls or numerical calculation. Furthermore, while hard constraints ensure exact physical consistency, soft constraints only achieve an approximation. In addition, soft constraints require more tuning effort due to the hyperparameter optimization needed for the regularization coefficient. On the other hand, they offer the flexibility to balance different terms when they are slightly contradictory. Table \ref{hard_vs_soft} summarizes these differences.

Soft constraints have dominated current research in seismology \citep{geng2025seismic,wang2026reciprocity}, owing to their ease of implementation. Making architectural changes needs significantly more effort. In this study, we propose RENO, a neural operator architecture that leverages a transformer backbone to enforce reciprocity as a hard constraint. The model obeys the imposed physical law by design, without the need for specialized training. Training is instead used to learn other aspects of wave propagation physics. The RENO inherits the full capabilities of conventional neural operators, while offering superior physical consistency and an order-of-magnitude computational speedup in multi-source scenarios. To start, the current architecture is designed for the reciprocity relation for particle velocity fields under single forces in the same direction, which is directly transferable to pressure fields under dilatational sources and travel-time fields governed by the eikonal equation. The framework supports various scientific applications, including ambient noise tomography, marine seismic surveys, and travel-time tomography. Future studies will address more complex reciprocity relations.

\begin{table}
\centering
\caption{Comparison of physics-informed methods in machine learning: hard and soft constraints.}
\label{hard_vs_soft}
\small
\newcolumntype{C}{>{\centering\arraybackslash}X} 

\begin{tabularx}{\textwidth}{@{}l >{\raggedright\arraybackslash}X C C C C@{}}
\toprule
\textbf{Constraint Type} & \textbf{Method} & \textbf{Training Efficiency} & \textbf{Physical Consistency} & \textbf{Tuning Effort} & \textbf{Flexibility} \\
\midrule
Hard & Embedded in Architecture & High & Exact & Low & Low \\ \addlinespace
Soft & Regularization & Low & Approximate & High & High \\
\bottomrule
\end{tabularx}
\end{table}

\section*{Data and Resources}
We have open-sourced our code at \href{https://github.com/caifeng-zou/RENO}{https://github.com/caifeng-zou/RENO}.

\section*{Acknowledgements}
This material is based upon work supported by the U.S. Department of Energy, Office of Science, Office of Advanced Scientific Computing Research, Science Foundations for Energy Earthshot under Award Number DE-SC0024705. This research is also partially supported by NSF-2438773.

\bibliography{main.bib}
\bibliographystyle{myst}

\label{lastpage}

\end{document}